\documentclass[modern]{aastex631}
\usepackage{graphicx}	
\usepackage{amsmath}	
\usepackage{float}	
\usepackage{nomencl}
\makenomenclature

\begin{document}

\title{Reflection from Inclined, Relativistic Light Sails}

\author[0000-0003-2790-8070]{Refath Bari}
\affiliation{The City College of New York \\
160 Convent Ave, New York, NY 10031}



\begin{abstract}
\cite{kipping2017relativistic} discusses relativistic reflection from light sails. As an extension to this work, we present a new formula for relativistic reflection of sunlight from a light sail moving at a velocity $v$ inclined at an arbitrary angle $\phi$, using an elementary two-body analysis. The formula is shown to reduce to the formulas of Euclid, Einstein, and Gjurchinovski in the relevant limits. This calculation may be extended to calculate the kinetic energy gained by a relativistic light sail inclined at an arbitrary angle.   
\end{abstract}


\section{Introduction}
The efficiency of chemical rocket propulsion is limited by the rocket equation \citep{benford2017sailships}. The search for alternative fuel sources, especially for interstellar travel, has been the subject of much recent exploration. Recent work has also explored the difficulty of exoplanetary civilizations using chemical rockets to escape super-Earths \citep{AviRocketChem,Hippke2018}. Electric sails and magnetic sails have been among the proposed alternatives which do not require any onboard fuel\citep{lingam2020electric}. Of particular interest have been photonic-based propulsion systems, which leverage blackbody radiation from stars or infrared emissions from planets to propel spacecraft to relativistic velocities \citep{santi2022swarm, lingam2017fast, gao2023dynamically, forgan2018photogravimagnetic, levchenko2018prospects, Carzana2022,Vulpetti2014}. Despite their appealing simplicity, light sails pose several engineering problems \citep{Spencer2019, santi2022multilayers}. \cite{Guillochon2015} have explored the effects of excess radiation leakage. The stability problems of light sails has also been extensively explored \citep{Manchester2017, savu2022structural, rafat2022self}. \cite{Atwater2018} have examined the material challenges of designing ultralight nanocraft suitable for interstellar travel. \cite{Campbell2021} have recently demonstrated that relativistic light sails must be significantly curved to maintain structural integrity at relativistic velocities. \cite{kipping2017relativistic} has already explored relativistic reflection from a light sail, but without accounting for the inclination of the sail. Inclined mirrors introduce non-trivial effects such as the Jaffe reflection limit \citep{mirzanejhad2016general}, which are relevant to the structural geometry and design of future nanocraft. In view of the aforementioned work demonstrating that light sails must billow and the need for a formula of relativistic reflection generalized to inclined sails, we present a formula for relativistic reflection from an inclined lightsail. We assume an idealized light sail with a perfectly reflecting mirror. Our result may be extended to calculate radiation pressure on a relativistic, inclined sail.

\section{Relativistic Reflection}
How does light reflect from a mirror? Simple: $\theta_i=\theta_r$. But how does light reflect from a mirror that is \textit{moving} at a constant velocity $v$? This is the well-known moving mirror problem, and it was originally solved by Einstein in his famous 1905 paper on Special Relativity \citep{15}. However, what if the mirror is \textit{inclined} at an angle $\theta$? This problem was recently solved by Gjurchinovski, starting from classical assumptions \citep{9}. In this paper, we introduce a new formula for relativistic reflection from an inclined mirror, derived from fully relativistic assumptions. Our formula is shown to reduce to three well-known relations: In the limit that the mirror is vertical ($\theta=90^{\circ}$), our equation reduces to Einstein's formula \citep{15}. In the limit that the mirror is stationary ($v=0$), our equation reduces to Euclid's formula \citep{maxOptics}. Lastly, our equation is in full agreement with Gjurchinovski's formula for inclined mirrors for $0^{\circ}\leq \theta \leq 90^{\circ}$\citep{2}.

We now supply a brief literature review of the moving mirrors problem. It is well known that the first analysis of light reflecting from a moving mirror was made in 1905 by Einstein in his famous Special Relativity paper \citep{15}, where he used Lorentz Transformations to develop the relativistic law of reflection. Subsequent developments include Bolotovskii and Stolyarov's investigation of Snell's Law and the Law of Reflection for moving mediums using Maxwell's Equations \citep{3}. More recently, Gjurchinovski has investigated reflections off a moving mirror at normal and oblique incidence, but starting from non-relativistic assumptions and using a motivated ansatz instead of a full derivation \citep{2,7,9}. Galli and Amiri have considered the problem from the perspective of the photon's exchange of momentum with the mirror \citep{1,12}. Maesumi recently demonstrated that reflection of a moving planar mirror is equivalent to reflection off a reflection from a stationary hyperbolic mirror, by considering a fan of rays emitted from a fixed light source \citep{8}.

We now present an equation for the relativistic law of reflection for a moving mirror inclined at an angle $\theta$. This is the first derivation of its form, to the author's best knowledge. A similar approach has been employed by \cite{25}, but starting with non-relativistic assumptions, without an inclined mirror, and without leveraging Hamilton's optomechanical analogy \citep{34}.
\begin{figure}
\centering
\includegraphics[width=16cm]{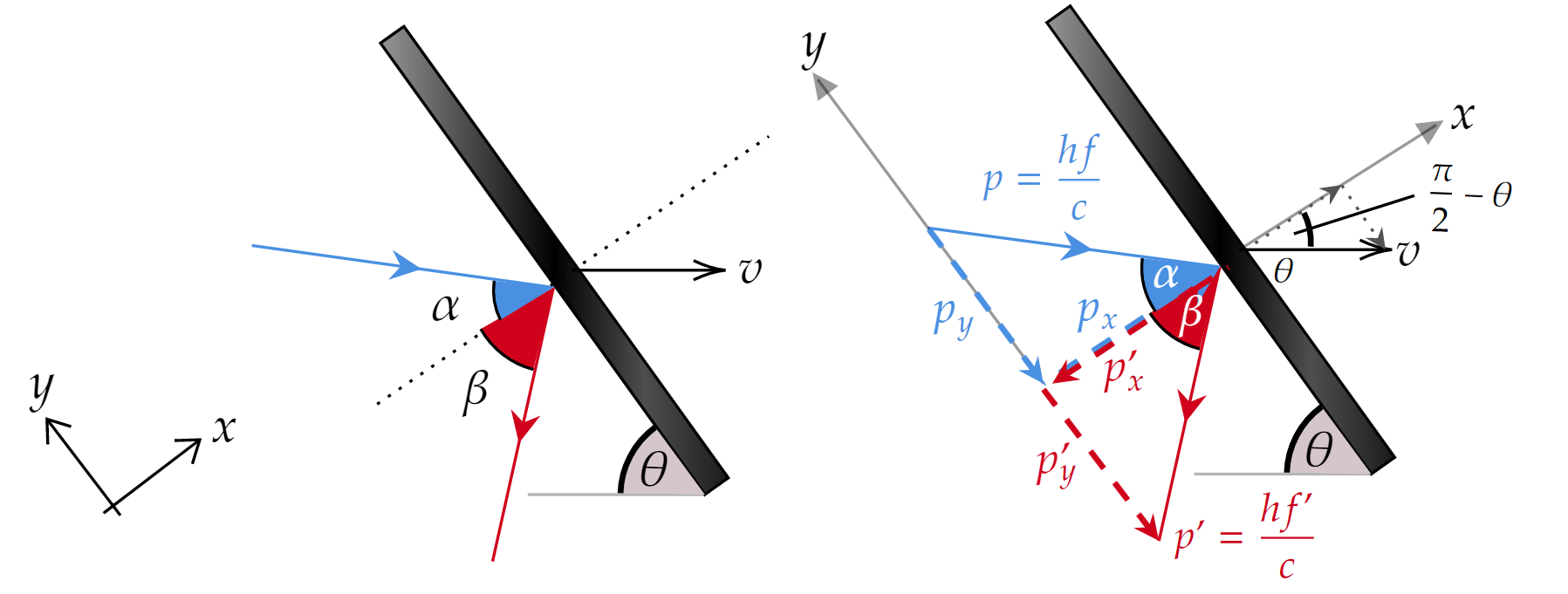}
\caption{Analysis of Reflection from an Inclined, Relativistic Light Sail}
\end{figure}
We have rotated our coordinate system such that the $x$ axis is along the normal to the mirror, and the $y$ axis is along the plane of the mirror. The initial and final momenta of the beam of light is given by $E=pc$:
\begin{gather*}
\text{Light}
\begin{cases}
  p_x: \frac{hf}{c}\cos{\alpha} \\
  p_y: -\frac{hf}{c}\sin{\alpha}
\end{cases}
\begin{cases}
  p'_x: -\frac{hf'}{c}\cos{\beta} \\
  p'_y: -\frac{hf'}{c}\sin{\beta}
\end{cases}
\end{gather*}
Likewise for the mirror of mass $M$ traveling at a constant velocity $v$, inclined at an angle $\theta$: 
\begin{gather*}
\text{Mirror}
\begin{cases}
  P_x: Mv\gamma \sin\theta \\
  P_y: -Mv\gamma \cos\theta
\end{cases}
\begin{cases}
  P'_x: M'v'\gamma' \sin\theta' \\
  P'_y: -M'v'\gamma' \cos\theta'
\end{cases}
\end{gather*}
By the conservation of momentum and energy, we immediately have
\begin{align}
  &\frac{hf}{c}\cos{\alpha} + Mv\gamma \sin\theta = -\frac{hf'}{c}\cos{\beta} + M'v'\gamma' \sin\theta' \label{eq:HorMomentumConservation}
  \\
  &
  -\frac{hf}{c}\sin{\alpha} -Mv\gamma \cos\theta = -\frac{hf'}{c}\sin{\beta} -M'v'\gamma' \cos\theta' \label{eq:2}
  \\
  &hf + \sqrt{(Mc^2)^2+(Pc)^2} = 
     hf' + \sqrt{(M'c^2)^2+(P'c)^2}
   \label{eq:EnergyConservation}
\end{align}
A billiard ball reflects as $\theta_i=\theta_r$ at a boundary precisely because the force exerted by the billiard table is normal to its surface. By the optomechanical analogy \citep{34}, we assume the contact force between the photon and mirror is also along the normal of the mirror, just as in the case for a billiard (i.e., no tangential forces). This implies that the vertical momentum of each the photon and the mirror remains invariant before and after the collision. Thus, $p_y=p'_y$ for the photon 
\begin{align}
  -\frac{hf}{c}\sin\alpha = -\frac{hf'}{c}\sin\beta \rightarrow f' = f \cdot \frac{\sin\alpha}{\sin\beta}
 \label{eq:firstFreq}
\end{align}
Likewise for the mirror, $P_y=P'_y$ by virtue of the contact force being only along the normal of the mirror. This gives an additional constraint:
\begin{align}
  -Mv\gamma \cos\theta = -M'v'\gamma' \cos\theta'
 \label{eq:optoMechAgain}
\end{align}
We seek an equation expressing the reflected angle as a function of the incident angle, velocity, and angle of inclination of the mirror: $\beta = \beta(\alpha, v, \theta)$. We will use \eqref{eq:EnergyConservation} to express $f'$ in terms of $f$. Setting this equal to \eqref{eq:firstFreq} will lead to cancelling $f'$ from both sides, resulting in the desired relation for $\beta$. We begin by rewriting \eqref{eq:EnergyConservation} in terms of the relativistic kinetic energy.
\begin{equation}
hf = hf' + \sqrt{(M'c^2)^2+(M'v'\gamma'c)^2} -Mc^2\gamma\label{eq:1}
\end{equation}
From conservation of momentum in the horizontal direction \eqref{eq:HorMomentumConservation}, we have
$$ \frac{hf}{c}\cos{\alpha} + Mv\gamma \sin\theta = -\frac{hf'}{c}\cos{\beta} + M'v'\gamma' \sin\theta'
$$
Letting $\mathcal{K}=(f\cos\alpha+f'\cos\beta)$, the final energy of the mirror is 
\begin{equation}
M'v'\gamma'c=\frac{h}{\sin\theta'}(f\cos\alpha+f'\cos\beta)+Mv\gamma c \frac{\sin\theta}{\sin\theta'}\label{eq:2}
\end{equation}
\begin{equation}
M'c^2=\frac{c}{v'\gamma'}\left\{\frac{h}{\sin\theta'}\mathcal{K}+Mv\gamma c \frac{\sin\theta}{\sin\theta'} \right\}\label{eq:3}
\end{equation}
Substituting \eqref{eq:2} and \eqref{eq:3} into \eqref{eq:1}, we find
\begin{equation}
hf = hf' + \left(\frac{h}{\sin\theta'}\mathcal{K} + Mv\gamma c \frac{\sin\theta}{\sin\theta'}\right)\sqrt{\left(\frac{c}{v'\gamma'}\right)^2 +1}-Mc^2\gamma\label{eq:4}
\end{equation}
Where $v'=v+\Delta v$. Our aim is now to simplify that second term. 
$$\sqrt{\left(\frac{c}{v'\gamma'} \right)^2 + 1}=\frac{c}{v+\Delta v}$$
Going back to our equation for the conservation of energy \eqref{eq:4}, we may now write
\begin{equation}hf = hf' + \left(\frac{h}{\sin\theta'}\mathcal{K} + Mv\gamma c \frac{\sin\theta}{\sin\theta'}\right)\frac{c}{v+\Delta v}-Mc^2\gamma\label{eq:5}\end{equation}
\begin{equation}
hf = hf' + \frac{hc}{\sin\theta'(v+\Delta v)}\mathcal{K}+Mc^2\gamma \left[\frac{v}{v+\Delta v}\left(\frac{\sin\theta}{\sin\theta'}\right)-1  \right] \label{eq:6}    
\end{equation}

We now substitute $\mathcal{K}$ into \eqref{eq:6}:
\begin{align*}
hf = hf' + \frac{hc}{\sin\theta'(v+\Delta v)}(f \cos \alpha + f' \cos \beta)+Mc^2\gamma \left[\frac{v}{v+\Delta v}\left(\frac{\sin\theta}{\sin\theta'}\right)-1  \right]
\end{align*}
Distributing out the right hand side, we have
\begin{align*}
    hf = hf' + \frac{hc}{\sin\theta'(v+\Delta v)}f\cos\alpha+\frac{hc}{\sin\theta'(v+\Delta v)}f'\cos\beta+Mc^2\gamma \left[\frac{v}{v+\Delta v}\left(\frac{\sin\theta}{\sin\theta'}\right)-1  \right]
\end{align*}
Separating the $f'$ and $f$ terms, we have: 
\begin{equation*}
\begin{split}
  f' = \frac{1}{1+\frac{c}{v+\Delta v}\frac{\cos\beta}{\sin\theta'}} \Bigl\{\, f\left(1-\frac{c}{v+\Delta v}\frac{\cos\alpha}{\sin\theta'}\right) - \frac{1}{h}Mc^2\gamma \left[\frac{v}{v+\Delta v}\left(\frac{\sin\theta}{\sin\theta'}\right)-1  \right]\,\Bigr\} \label{eq:7}
\end{split}
\end{equation*}

As per \cite{25}'s argument for relativistic reflection from a moving mirror, we assume that $M=M'$. Thus, the conservation of momentum in the horizontal direction (\ref{eq:2}) becomes
\begin{equation}
     Mc = \frac{h\mathcal{K}}{v'\gamma'\sin\theta'-v\gamma\sin\theta} \label{trickEq}
\end{equation}
We may thus rewrite $f'$ as
$$
  f' = \frac{1}{1+\frac{c}{v+\Delta v}\frac{\cos\beta}{\sin\theta'}} \Bigl\{\, f\left(1-\frac{c}{v+\Delta v}\frac{\cos\alpha}{\sin\theta'}\right) - \frac{1}{h}\frac{(Mc)^2}{M}\gamma \left[\frac{v}{v+\Delta v}\left(\frac{\sin\theta}{\sin\theta'}\right)-1  \right]\,\Bigr\}
$$
Substituting \eqref{trickEq} into the above equation, we find
\begin{equation}
\begin{split}
  f' = \frac{1}{1+\frac{c}{v+\Delta v}\frac{\cos\beta}{\sin\theta'}} \Bigl\{\, &f\left(1-\frac{c}{v+\Delta v}\frac{\cos\alpha}{\sin\theta'}\right) - \\ &\frac{1}{h}\frac{1}{M}\left[\frac{h\mathcal{K}}{v'\gamma'\sin\theta'-v\gamma\sin\theta}\right]^2 \gamma \left[\frac{v}{v+\Delta v}\left(\frac{\sin\theta}{\sin\theta'}\right)-1  \right]\,\Bigr\}
\end{split}
\end{equation}
In the limit that the mirror's mass is infinite, we may drop the last term \citep{25}. This is a reasonable approximation since the mass of the mirror is 'infinitely' larger than the massless photon. We have thus made no simplifications or assumptions beyond the validity of the conservation of energy and momentum. Since $\lim_{\Delta v\to 0} (\gamma'v'-\gamma v)=\gamma \Delta v /c$, which cancels the $\Delta v$ in the numerator as $M \to \infty$:
\begin{equation}
f'=f \left(\frac{1-\frac{c}{v}\frac{\cos \alpha}{\sin \theta'}} {1+\frac{c}{v}\frac{\cos \beta}{\sin \theta'}}\right) \label{eq:secondFreq}
\end{equation}
To eliminate $\theta'$, we return to the second equation offered by the optomechanical analogy, \eqref{eq:optoMechAgain}. Recall our assumption that $M=M'$ above. Furthermore, in the limit that $M\rightarrow \infty$, $\Delta v \rightarrow 0$ so that we are forced to conclude that 
\begin{equation}
\cos\theta'=\cos\theta \rightarrow \theta'=\theta \rightarrow f'=f \left(\frac{1-\frac{c}{v}\frac{\cos \alpha}{\sin \theta}} {1+\frac{c}{v}\frac{\cos \beta}{\sin \theta}}\right) \label{eq:finalFreq}
\end{equation}

We now set our two relations for $f'$ equal, \eqref{eq:firstFreq} and \eqref{eq:finalFreq}, which gives 
\begin{equation}
    \left(\frac{v\sin\theta-c\cos \alpha} {v\sin\theta+c\cos \beta}\right) = \frac{\sin\alpha}{\sin\beta} \label{eq:almostFinal}
\end{equation}
This is the desired relation: \eqref{eq:almostFinal} is the relativistic reflection law for a moving mirror inclined at an angle $\theta$. However, we must now perform three sanity checks: 
\begin{enumerate}
    \item Between $0\leq \theta \leq 90$ (horizontal and vertical mirror) and $0<v<c$, our formula should smoothly interpolate between the standard law of reflection and the relativistic law of reflection. Further, we expect it to conform to Gjurchinovski's relation for reflection off a moving, inclined mirror \cite{2}. $$\cos \beta=\frac{-2 \frac{(v\sin\theta)}{c}+\left(1+\frac{(v\sin\theta)^2}{c^2}\right) \cos \alpha}{1-2 \frac{(v\sin\theta)}{c} \cos \alpha+\frac{(v\sin\theta)^2}{c^2}}$$
    \item In the upper limit that $\theta=90^{\circ}$, we have a vertical moving mirror and our equation should reduce to Einstein's original relativistic reflection formula \cite{15}.
    \item In the lower limit that $v=0$, the mirror is not moving and our equation should reduce to Euclid's Law of Reflection \cite{maxOptics}.
\end{enumerate}
\begin{figure}\label{BariGjur}
\centering
\includegraphics[width=11cm]{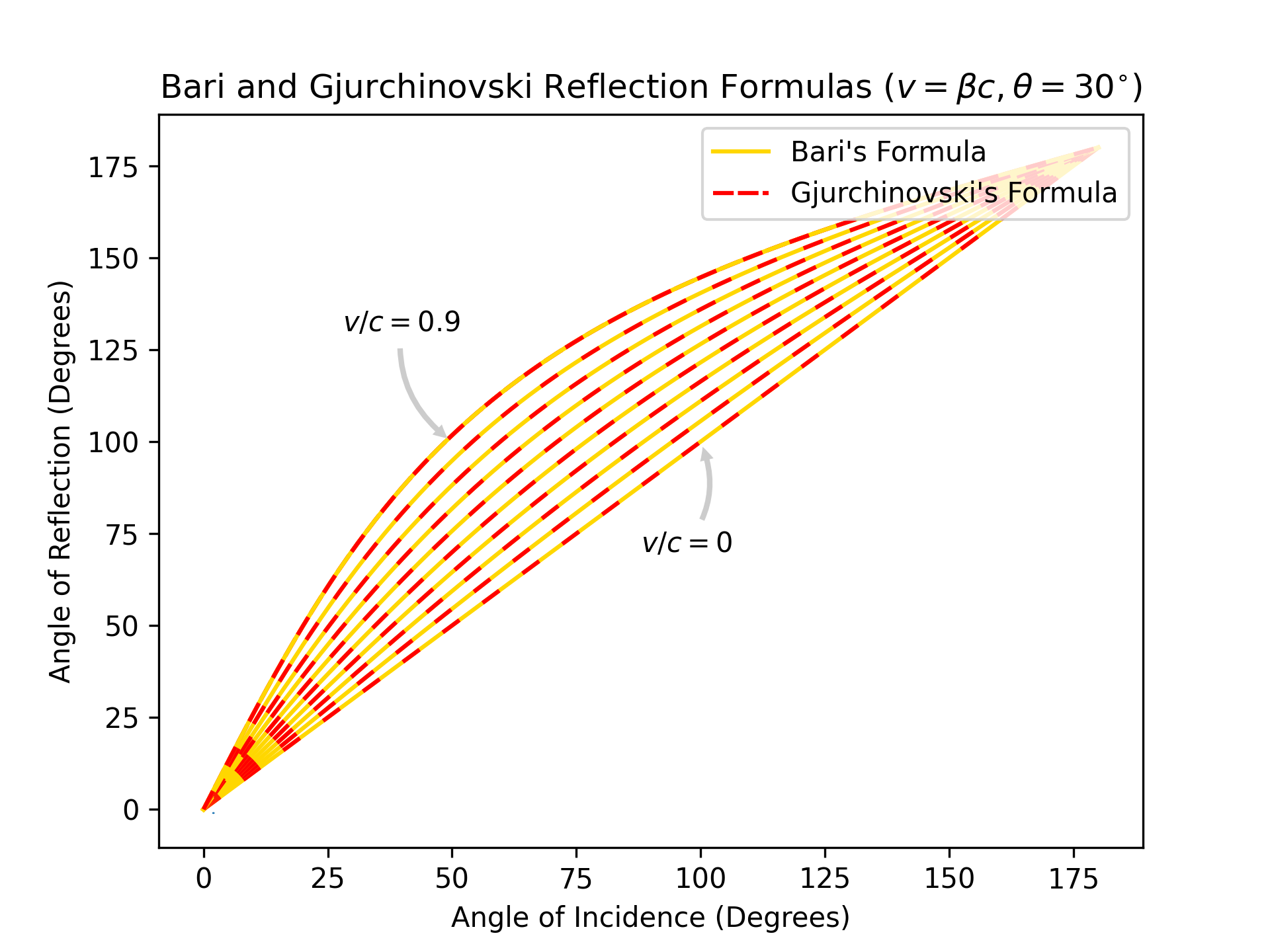} 
\caption{Eq. \eqref{eq:BarisLaw} against Gjurchinovski's Formula; Lines drawn at $\beta = 0.0, 0.1, ..., 0.9$}\label{BariComparison}
\end{figure}

We begin by ensuring that the formula interpolates correctly between $0\leq\theta\leq90$, and agrees with Gjurchinovski's relation. Indeed, upon plotting them against each other, we find that they match exactly. The only caveat is that our formula seems to agree with the right side of Gjurchinovski's relation. In Gjurchinovski's equation, the horizontal component of the mirror's velocity is $v_x=v\cos\theta$, but in our setup, $v_x=v\sin\theta$. Upon observing the plot of \eqref{eq:almostFinal}, we observe that it delivers the angle of reflection for $\alpha>180$, which is manifestly unphysical. By negating our equation, we obtain the angle of reflection for $0<\alpha<90$, as desired. Our final relativistic reflection law for inclined mirrors is thus 
\begin{equation}
    -\left(\frac{v\sin\theta-c\cos \alpha} {v\sin\theta+c\cos \beta}\right) = \frac{\sin\alpha}{\sin\beta} \label{eq:9}
\end{equation}
It is worth solving explicitly for the angle of reflection $\beta$, so that our formula matches the format of Einstein's and Gjurchinovski's. Letting $k=(v/c)\sin\theta$, we have
\begin{equation}
        -\left(\frac{(v/c)\sin\theta-\cos \alpha} {(v/c)\sin\theta+\cos \beta}\right) = \frac{\sin\alpha}{\sin\beta} \rightarrow     
        \frac{\cos\alpha-k}{\cos\beta+k}=\frac{\sin\alpha}{\sin\beta}
 \label{eq:9}
\end{equation}
$$ \sin(\beta-\alpha) = k(\sin \alpha + \sin \beta)=2k \sin \frac{\alpha+\beta}{2}\cos \frac{\beta-\alpha}{2} $$
Expanding by trigonometric relations, we have
$$ 2\sin \frac{\beta-\alpha}{2}\cos \frac{\beta-\alpha}{2} - 2k \sin \frac{\alpha+\beta}{2}\cos \frac{\beta-\alpha}{2} =0$$

Simplifying, we isolate the reflected angle $\beta$ to obtain
$$(k-1)\tan \frac{\beta}{2} +(k+1)\tan \frac{\alpha}{2}=0$$

\begin{figure}[h]\label{BariEinstein}
\centering
\includegraphics[width=11cm]{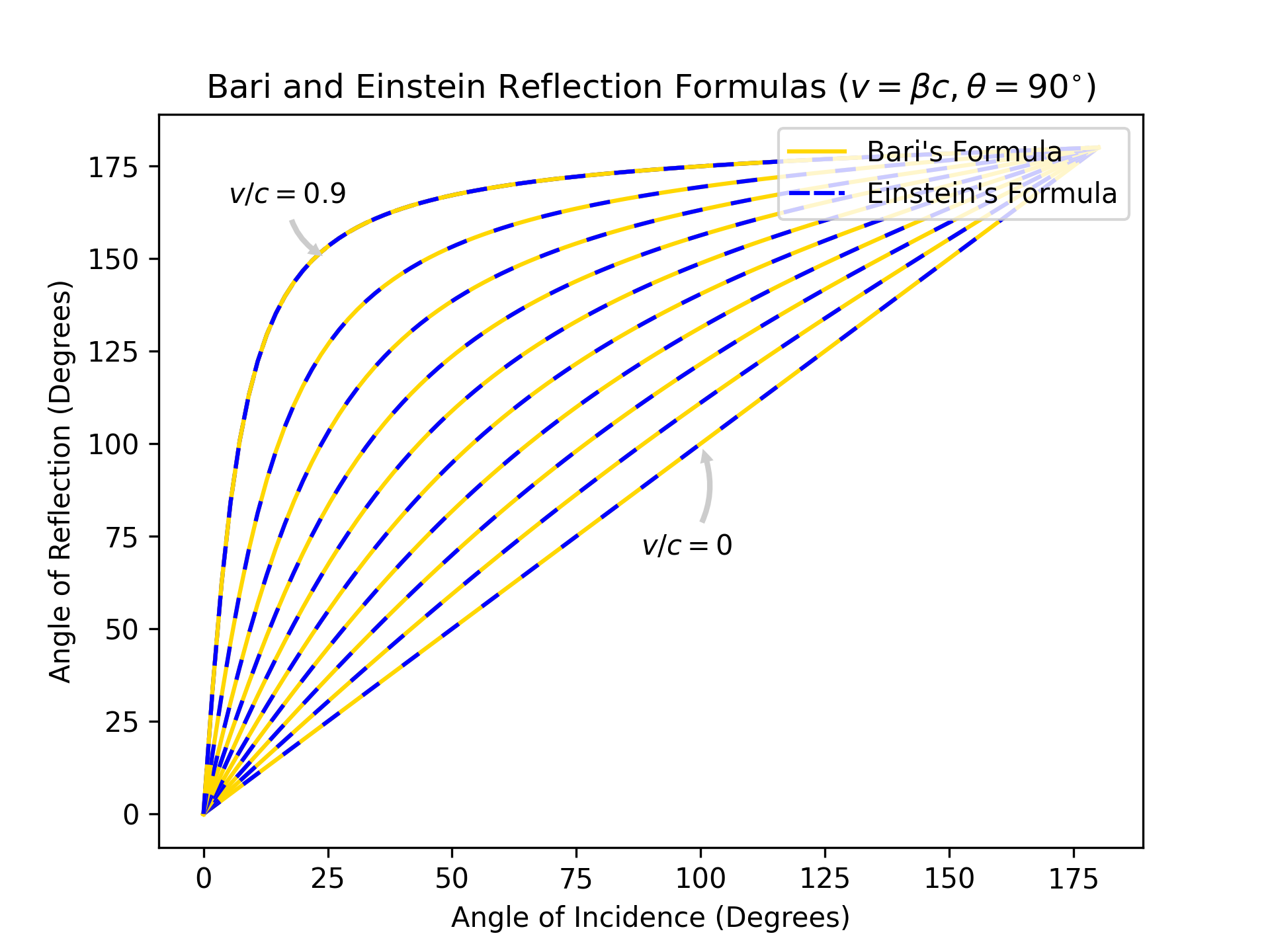} 
\caption{Eq. \eqref{eq:BarisLaw} against Einstein's Formula; Lines drawn at $\beta = 0.0, 0.1, ..., 0.9$}\label{BariComparison}
\end{figure}

\eqref{eq:BarisLaw} is our final result. We have shown that it agrees with Gjurchinovski. We now proceed to the second check, that it reduces to Einstein's Formula for a vertical mirror when $\phi=90$. As shown by \ref{BariEinstein}, we observe exact numerical agreement.

\begin{equation}
    \boxed{\beta = 2\arctan\left(-\frac{(v/c)\sin\theta+1}{(v/c)\sin\theta-1}\tan\left(\frac{\alpha}{2}\right)\right)}\label{eq:BarisLaw}
\end{equation}

Finally, for our third check, does \eqref{eq:BarisLaw} reduce to Euclid's Law of Reflection for $v=0$? In that case, 
$$\beta = 2\arctan\left(\tan\left(\frac{\alpha}{2}\right) \right) \rightarrow \alpha = \beta$$
We have thus demonstrated that \eqref{eq:BarisLaw} agrees with the formulas of Euclid, Einstein, and Gjurchinovski. Furthermore, we have derived this formula from basic relativistic assumptions, in conjunction with the opto-mechanical analogy. We thus have a high confidence in the fidelity of our derived relation.

\section{Discussion}
In this paper, a relativistic law of reflection for an inclined solar sail was developed by considering the sail-photon system as a two-body problem. The objective was to use basic physical principles to determine the angle of reflection $\beta$ of a photon given an incident angle $\alpha$ and a solar sail of mass $M$, velocity $v$, and inclination angle $\theta$. The conservation of energy provided one equation and the conservation of momentum provided two equations (one for each direction). However, this was not sufficient: we still could not find $\theta'$ in terms of $\theta$, for instance. To advance forward, we exploit Hamilton's optomechanical analogy, which treats the photon as a billiard ball, exchanging contact force with the mirror only along its perpendicular. This provides two additional equations which ultimately enabled us to determine $\beta$.

We now note an alternative approach to this derivation. In this paper, we approached the problem of radiation from a relativistic, inclined light sail as a two-body problem employing the conservation of energy, momentum, and the optomechanical analogy. An alternative approach would be to transform to the frame of the moving light sail, in which the standard law of reflection $\theta_i = \theta_r$ applies. Upon returning to the frame of reference in which the sail is moving, we employ the lorentz transformation on the standard law of reflection to obtain its relativistic counterpart. In fact, this was exactly the strategy employed by \cite{15} in his famous special relativity paper.

Our assumption that the mirror is of "infinite" mass $M\rightarrow \infty$ so that $Mc^2>>hf$ implies that there is no acceleration of the sail. That would make the entire venture of solar sailing a moot point. The caveat is that an ensemble of many photons is required to result in the acceleration of a solar sail. It should also be noted that photon must hit the center of the mirror, otherwise the mirror will begin rotate, by virtue of the conservation of angular momentum. The photon, despite being massless, not only carries a linear momentum $\hbar k$, but also an angular momentum $\hbar$. 

It is straightforward to obtain the redshift of the reflected photon from \eqref{eq:finalFreq}. One may therefore calculate the kinetic energy gained by the light sail by equating it to the total energy lost by the photon, which would simply be the amount by which it is redshifted. We also note that above a certain incident angle, given by the Jaffe Reflection Limit \citep{jaffe1973forward}, the photon will no longer reflect off the mirror. This interesting phenomenon occurs when the horizontal component of the mirror's velocity exceeds the horizontal component of the photon's velocity. The photon thus never "reaches" the mirror and thus never reflects. Transforming to the mirror's frame of reference, this corresponds to a photon "incident" at $\theta_i>90^{\circ}$ \citep{7}. This phenomenon may be of importance for the geometry of future relativistic solar sail designs.

Our result may also be of relevance to interstellar solar sails, for which a minor deviation (i.e., one percent error) can result in the sail missing the intended target by a significant amount. Recently, solar thermal propulsion (STP) systems have been explored as a potential means of propelling microsatellites on interstellar missions. STP systems exploit mirrors and lenses to focus solar radiation to heat propellants, resulting in a higher specific impulse and exhaust velocity than cold gas thrusters, which are the conventional propulsion systems for small satellites. Future solar sails that are moving at a fraction of the speed of light may be able to exploit the relativistic law of reflection to focus solar radiation to heat propellants. Indeed, the construction of a relativistic STP system would hinge on Eq. \eqref{eq:BarisLaw}. To construct such a mirror for a relativistic STP system, one may take inspiration from \cite{8}, who demonstrated that relativistic reflection from a vertical, plane mirror is equivalent to standard reflection from a certain hyperbolic mirror. Similarly, one may consider a fan of light rays approaching a curved mirror, and apply \eqref{eq:BarisLaw} at each infinitesimally small segment of the mirror to determine what mirror geometry would result in focusing solar radiation.  
\section{Appendix}
We now briefly introduce the notation which is used above in the derivation: 
\nomenclature{$\alpha$}{Photon's angle of incidence against solar sail}
\nomenclature{$\beta$}{Photon's angle of reflection against solar sail; This symbol is used in a different context as $\beta=v/c$ in the titles of Figures 2 and 3}
\nomenclature{$\theta$}{Angle of inclination of solar sail}
\nomenclature{$M$}{Mass of solar sail}
\nomenclature{$v$}{Speed of solar sail}
\nomenclature{$\gamma$}{Lorentz gamma factor}
\nomenclature{$p_i, p'_i$}{Initial and final momentum for the photon in the $\hat{i}$ direction (i.e., $\hat{x},\hat{y}$)}
\nomenclature{$P_i, P'_i$}{Initial and final momentum for the mirror in the $\hat{i}$ direction (i.e., $\hat{x},\hat{y}$)}

\printnomenclature



\bibliographystyle{aasjournal}
\bibliography{References} 



\end{document}